# Gene targeting in disease networks

**Deborah Weighill[1], Marouen Ben Guebila[1], Kimberly Glass[1,2,3], John Platig[2,3], Jen Jen Yeh[4] and John Quackenbush[1,2,*]**

[1]Department of Biostatistics, Harvard T.H. Chan School of Public Health, Harvard University, Boston, MA, USA

[2]Channing Division of Network Medicine, Brigham and Women's Hospital, Boston, MA, USA

[3]Harvard Medical School, Harvard University, Boston, MA, USA

[4]Departments of Surgery and Pharmacology, Lineberger Comprehensive Cancer Center, University of North Carolina at Chapel Hill, Chapel Hill, NC, USA

**\* Correspondence:**
John Quackenbush
johnq@hsph.harvard.edu

**Abstract**

Profiling of whole transcriptomes has become a cornerstone of molecular biology and an invaluable tool for the characterization of clinical phenotypes and the identification of disease subtypes. Analyses of these data are becoming ever more sophisticated as we move beyond simple comparisons to consider networks of higher-order interactions and associations. Gene regulatory networks model the regulatory relationships of transcription factors and genes and have allowed the identification of differentially regulated processes in disease systems. In this perspective we discuss gene targeting scores, which measure changes in inferred regulatory network interactions, and their use in identifying disease-relevant processes. In addition, we present an example analysis or pancreatic ductal adenocarcinoma demonstrating the power of gene targeting scores to identify differential processes between complex phenotypes; processes which would have been missed by only performing differential expression analysis. This example demonstrates that gene targeting scores are an invaluable addition to gene expression analysis in the characterization of diseases and other complex phenotypes.

## 1    Introduction

A core tenet of molecular biology is that phenotypic differences are reflected through patterns of differential expression of key genes involved in relevant biological processes. Since its inception, whole-genome transcript profiling been an invaluable tool for exploring these associations and have been used in a range of applications, including identification of clinically relevant molecular subtypes in cancer exhibiting different morbidities and implications for treatment together with characteristic genes associated with these phenotypes (Moffitt et al. 2015; Rouzier et al. 2005; Rudin et al. 2019; Bailey et al. 2016; Rodriguez-Salas et al. 2017; Collisson et al. 2011; Kwa, Makris, and Esteva 2017; Sjödahl et al. 2019). Studies have also found that complex patterns of association between gene expression, represented as networks, can provide additional insight and that network metrics that parameterize these associations can be used to prioritize and identify crucial disease-related genes (Ramadan, Alinsaif, and Hassan 2016; Dimitrakopoulos et al. 2018; Gumpinger et al. 2020; Horn et al. 2018). However, there is growing evidence that the processes regulating the expression of phenotype-associated genes can provide a more holistic picture of drivers of disease and other phenotypes. Gene regulatory networks (GRNs are often represented as directed bipartite graphs that are used to depict inferred relationships between transcription factors (TFs) and their target genes. GRNs can be characterized by calculating the "gene targeting score," a network topology measure that captures the complex relationships that a gene has with other TFs and genes and represents the extent to which a gene is targeted in a given system. In this perspective, we will present gene targeting scores, discuss their meaning, and show how this network-based measure provides information about disease systems beyond that found using only differential expression analysis in the investigation and characterization of human disease.

## 2    Gene regulatory networks: characterization of systems

Networks are useful tools for representing and analyzing large, complex datasets because they capture information about the *relationships* within a system rather than simply the state of individual components. This is an important distinction, as can be illustrated with a small toy example first described in (Glass et al. 2014) (Figure 1). In this example, we consider the expression of four genes in nine healthy individuals and nine individuals with a disease (Figures 1A and 1B). Comparing expression levels between healthy and diseased individuals, we find that none are differentially expressed (Figures 1C and 1D). However, when looking at the co-expression of these genes in each group of nine, we see that that the genes are *differentially co-expressed* between groups. For example, in healthy individuals, gene G1 is co-expressed with gene G2 (Figure 1E), whereas in diseased individuals, gene G1 is co-expressed with gene G3 (Figure 1F). This illustrates that differential expression analysis alone may miss important correlations or regulatory relationships that distinguish biological states such as healthy and disease.

This does not mean that gene expression analysis alone is not useful. Differential expression analyses have contributed to many key advances in our understanding of disease. For example, much of our understanding of the complexities of human cancers is derived from large scale expression profiling of cancer, such as that carried out by The Cancer Genome Atlas (TCGA; https://www.cancer.gov/tcga), where the expression-based subtypes that have been identified possess distinct clinical characteristics. In pancreatic ductal adenocarcinoma (PDAC), several studies have used expression profiling to refine molecular subtypes (PDAC; Moffitt et al. 2015; Rashid et al. 2020; Collisson et al. 2011; Bailey et al. 2016 Puleo ; Maurer). However, we suggest that a more comprehensive molecular characterization of diseases can be achieved by exploring inferred regulatory network differences and differential gene targeting.

In 2013, Glass *et al.* introduced the PANDA (Passing Attributes between Networks for Data Assimilation) framework for GRN construction (Glass et al. 2013). This method takes a unique approach to GRN construction by using message passing to integrate multiple data sources. PANDA predicts regulatory relationships between TFs and genes by considering three main sources of information: (1) a TF-gene network "adjacency matrix" representing an initial guess of which TFs regulate which genes based on the presence/absence of a TF motif in the promoter region of the gene, (2) a protein-protein interaction network "cooperativity matrix" that recognizes that many TFs exert their influence through regulatory complexes, and (3) a gene co-expression network matrix representing gene-gene relationships initially based on correlation in expression patterns across a set of samples. These three different sources of information are iteratively updated using a message-passing algorithm, using the logic that if two genes are co-expressed, they are more likely to be co-regulated by a similar set of TFs (Figure 2A) and that if two TFs interact, they are more likely to bind promoter regions as a complex and co-regulate the expression of their target genes (Figure 2B). In this process, the TF-gene "edge weights" in the adjacency matrix are updated to reflect the evidence supporting a regulatory interaction; the refinement of edge weights through message passing has been found to improve the prediction accuracy of GRNs, validated through prediction of ChIP-seq binding.

PANDA and has been used to investigate gene regulatory relationships in several disease contexts, including chronic obstructive pulmonary disease (Glass et al. 2014), asthma (Qiu et al. 2018), ovarian cancer (Glass et al. 2015), and colorectal cancer (Lopes-Ramos et al. 2018; Vargas, Quackenbush, and Glass 2016). In addition, single-sample versions of PANDA GRNs, derived using a method called LIONESS (Linear Interpolation to Obtain Network Estimates for Single Samples) (Marieke Lydia Kuijjer et al. 2019), have been used to study sex-linked differences in colon cancer (Lopes-Ramos et al. 2018) as well sex-related differences in gene regulation in tissues (Lopes-Ramos et al. 2020).

## 3    Gene targeting score: identifying informative regulatory processes

The use of GRNs in the analysis of disease relies on analysis of the "gene targeting score," a numerical score representing the extent to which a gene is targeted by TFs in a given biological context. The gene targeting score is calculated by summing the weights of all inbound regulatory edges for a gene (Figure 2C). Because of the way in which PANDA estimates edge weights, a gene's targeting score synthesizes multiple lines of evidence—TF motif data, TF-TF interactions, and expression correlation. Thus gene targeting scores are not necessarily correlated with absolute gene expression levels and consequently differential targeting is not necessarily correlated with differential gene expression.

Sonawane et al. (2017) used PANDA to construct tissue specific GRNs for 38 tissues in GTEx to investigate the tissue-specificity of TF-gene regulatory relationships. They found many tissue-specific regulatory relationships which would have been missed by using expression information alone. For example, when comparing tissue-specific regulatory activity of TFs based on gene expression to that deduced using network targeting, they found that TF regulation of tissue-specific function was evident when using gene targeting metrics but it was largely independent of TF expression level (Sonawane et al. 2017).

Glass et al. (2014) investigated sexual dimorphism in Chronic Obstructive Pulmonary Disease (COPD), a disease for which women have a higher susceptibility. They first compared gene expression between males and females with COPD and found little evidence of differential gene expression of autosomal genes between the sexes. They then constructed jack-knife ensembles of male and female GRNs using PANDA. After calculating the gene targeting score for each gene in each network, they found that many genes were significantly differentially targeted between males and females. Pre-ranked gene set enrichment analysis based on gene targeting identified several biological processes enriched for genes with higher targeting in females; these were all related to mitochondrial function, which has been previously been implicated in many aspects of COPD and lung disease (Glass et al. 2014).

Lopes-Ramos et al. (2020) investigated sex-differences in gene expression and gene regulation in twenty-nine human tissues by constructing individual-specific networks for each sample in each tissue. Differential edge weights between males and females were identified, and genes were classified as differentially targeted if at least 5% of their inbound edge weights were significantly different between males and females. This allowed genes to be classified as being male-biased, if most (>60%) of the inbound differential edges were higher in males, female-biased, if most (>60%) of the inbound differential edges were higher in females, and sex-divergent, if the number of inbound differential edges was similar between being higher in males and higher in females (Lopes-Ramos et al. 2020). Consistent with previous studies, they found little differential gene expression except in breast tissue, with the median number of differentially expressed genes across tissues equal to 64. However, wide-spread sex-biased targeting was detected in all tissues, with a median number of differentially targeted genes across tissues equal to 169. Interestingly, the sex hormone receptors ESR1, ESR2, and AR were differentially targeted between male and female individuals in several tissues such as breast, heart, and blood, despite the fact that those hormone receptors were not differentially expressed.

## 4     Specific example: Pancreatic ductal adenocarcinoma subtypes

PDAC is a lethal disease involving heterogenous tumors that comprised of diverse cell types including tumor epithelial cells and as components of the tumor microenvironment such as immune cells and fibroblasts. Molecular subtypes of PDAC have been identified through gene expression analysis (Collisson et al. 2011; Puleo et al. 2018; Moffitt et al. 2015; Bailey et al. 2016; Maurer et al. 2019) and the basal-like and classical subtypes have been associated with both prognosis and treatment response(Moffitt et al. 2015; Rashid et al. 2020; O'Kane et al. 2020; Aung et al. 2018).

To see what additional insight we could gain into these subtypes using network methods and differential targeting, we compared differential gene expression and differential GRN gene targeting scores between the networks derived from basal-like and classical subtypes of PDAC for 150 tumors from TCGA (Cancer Genome Atlas Research Network 2017) using the TPM (transcripts per kilobase million) expression data from Recount (Ettou et al. 2020). We used PANDA and LIONESS to construct sample-specific GRNs and chose to chose to limit our analysis to those genes with a high standard deviation of logTPMs (sd(logTPMs) > 0.4) across samples. For each gene in each individual tumor, a gene targeting score was calculated as the sum of all inbound edges surrounding the gene. Separately, we also calculated a sample-specific co-expression network for each tumor (Marieke L. Kuijjer et al. 2019) and for each gene in each sample, calculated a gene co-expression score equal to the sum of each gene's co-expression edge surrounding the gene. We used limma (Ritchie et al. 2015) to compare the expression data, the correlation networks, and GRNs between the basal-like and classical subtypes, allowing us to identify differentially expressed genes, differentially co-expressed genes, and differentially targeted genes, respectively.

The three genes found to be most significantly differentially targeted in GRNs, but not differentially expressed, between basal-like and classical subtypes are folate receptor beta (FOLR2), hedgehog interacting protein (HHIP), and the CD209 antigen C-type lectin domain family 4 member L (CD209). FOLR2 encodes the folate receptor 2 protein and is known to be overexpressed in tumor-associated macrophages (Tie et al. 2020). HHIP codes for the hedgehog interacting protein; the hedgehog signaling pathway regulates cell differentiation and proliferation and is activated in several cancers including PDAC (Yang et al. 2010; Gu, Schlotman, and Xie 2016; Honselmann et al. 2015). CD209 codes for a C-Type lectin domain family 4 protein, and is a dendritic cell marker. The roles that these play in PCAC have not yet been explored.

For each of the ranked lists of differentially targeted genes, differentially co-expressed genes, and differentially expressed genes, we performed pre-ranked gene set enrichment analysis (Eden et al. 2009; Supek et al. 2011) to identify significantly over-represented functional gene sets. Both the differential targeting analysis derived from GRNs and differential expression analysis identified keratinization, cornification, cell death, and wound healing as differentiating between basal-like and classical samples. However, several immune-related processes, epigenetic, and cell cycle process found by differential targeting analysis were missed using differentially expressed genes (Figure 2). Functional enrichment using co-expression scores to rank genes identified some processes similar to those found using differential targeting but missed several important pathways related to cell cycle and other significant processes, such as chromatin organization.

Genes encoding keratins and laminins are biomarkers for basal-like tumors (Moffitt et al. 2015). The fact that both differential expression and differential targeting identified keratinization and cell adhesion as biological processes distinguishing PDAC subtypes serves as an internal consistency check. Among the processes found only by differential targeting, processes related to the immune system speak to the importance of the tumor microenvironment, which is known to influence PDAC prognosis and drug response; a high degree of tumor-associated macrophage infiltration has been linked to lower survival (Karamitopoulou 2019). The differentially targeting of epigenetic functions between subtypes is also important is consistent with reports that PDAC subtypes have distinct epigenetic landscapes (Lomberk et al. 2018).

This analysis of PDAC subtypes, although abbreviated, demonstrates the power of using GRN inference and gene targeting score analysis to identify regulatory processes that characterize distinct phenotypes—including processes that are distinct from those that are associated with patterns of gene expression. The biologically relevant differences we see in targeting but not expression or co-expression suggest that regulatory control, even if not activated, is important in defining health and disease.

## 5     Discussion

There is growing experimental evidence of the importance of complex regulatory processes in distinguishing phenotypes in health and disease. For example, the Wilms tumor-1 (WT1) TF is a master regulator that targets several essential genes in kidney podocyte cells. Ettou et al. (2020) investigated WT1-based gene regulation during podocyte injury and observed an increase in binding intensity at WT1-bound regions, as well as an increase in expression of several hallmark podocyte genes. They found that WT1 maintained open chromatin in the regions of its target genes but that expression level of WT1 was not universally associated with the intensity of its binding. The WT1 binding pattern was found to be distinct at each binding site and they determined that regulation by WT1 could cause either an increase or a decrease in the expression of its target genes. Importantly, the genes regulated by WT1 are associated with biological processes associated with response to podocyte injury. The role of complex regulatory processes is further illustrated by the work of Carnesecchi et al. (2020), who investigated how a single transcription factor could regulate different developmental programs in various cell lineages. They showed that the Ubx transcription factor forms different complexes with distinct binding partners in various cell lineages despite the fact that most of the interaction partners showed no differential expression across the lineages.

Taken together, the results reported by Ettou and Carnesecchi illustrate the complexity of regulatory processes and the importance that regulatory targeting plays in defining phenotype, even in instances in which a key regulator does not itself substantially change its expression levels. Their results also point to the importance of modeling both "direct" and "indirect" regulation of genes by TFs and the complexes they form. Among methods for GRN inference, PANDA (and by extension, PANDA+LIONESS) is singular in considering interactions between TF proteins in its model, and thus may provide unique insight into regulatory processes. And PANDA's integrative approach to allows TF-TF interactions, predicted TF-gene regulatory, and gene co-expression data to be further refined to optimize agreement between these information sources, leading to better gene targeting predictions.

The work summarized in this perspective demonstrates the value of the gene targeting score as a metric for assessing the drivers of phenotypic differences. Gene targeting scores not only capture structural characteristics of regulatory networks, but also allow for the identification of processes that may be activated in response to appropriate stimuli and so help to define phenotypes and disease subtype. For example, the analysis of sex-linked differences in colon cancer drug metabolism (Lopes-Ramos et al. 2018) was performed using gene expression data obtained from surgical samples obtained. Giving that the samples had not been collected prior to the initiation of chemotherapy, is not surprising, then, that expression (or co-expression) analysis did not identify sex differences in drug metabolism. However, the discovery of regulatory differences through gene targeting analysis in genes involved drug metabolism and related processes suggests that male and female tumor cells are programmed to respond differently—and do so when treated with chemotherapy drugs, leading to the clinically observed differences in drug responses between males and females. Our application of PANDA and LIONESS in comparing PDAC subtypes shows that differential targeting analysis can provide new insights into disease-associated processes—insights missed when looking at expression or co-expression.

The PANDA and LIONESS tools necessary for GRN analyses and identification of differential targeting are freely available with extensive documentation (netzoo.github.io) and can easily be implemented in most analytical workflows. We hope that this review motivates the broader use and appreciation of gene targeting analysis.

## 6      Conflict of Interest

*The authors declare that the research was conducted in the absence of any commercial or financial relationships that could be construed as a potential conflict of interest.*

## 7      Author Contributions

All authors discussed, planned, reviewed and edited the manuscript. DW drafted the manuscript and performed the analysis.

## 8      Funding

DW, MBG, and JQ are supported by a grant from the US National Cancer Institute (NCI), R35CA220523; MBG and JQ are further supported by NCI grant U24CA231846. KG is supported by a grant from the US National Heart, Lung, and Blood Institute (NHLBI), K25HL133599. JP is supported by a grant from the US National Heart, Lung and Blood Institute, K25HL140186. JJY is supported by grant from the NCI: R01CA199064, CA193650 and U24CA211000.

## 9     Data Availability Statement

Publicly available data analyzed in this study was obtained from recount2 (https://jhubiostatistics.shinyapps.io/recount/).

## 11 Figures

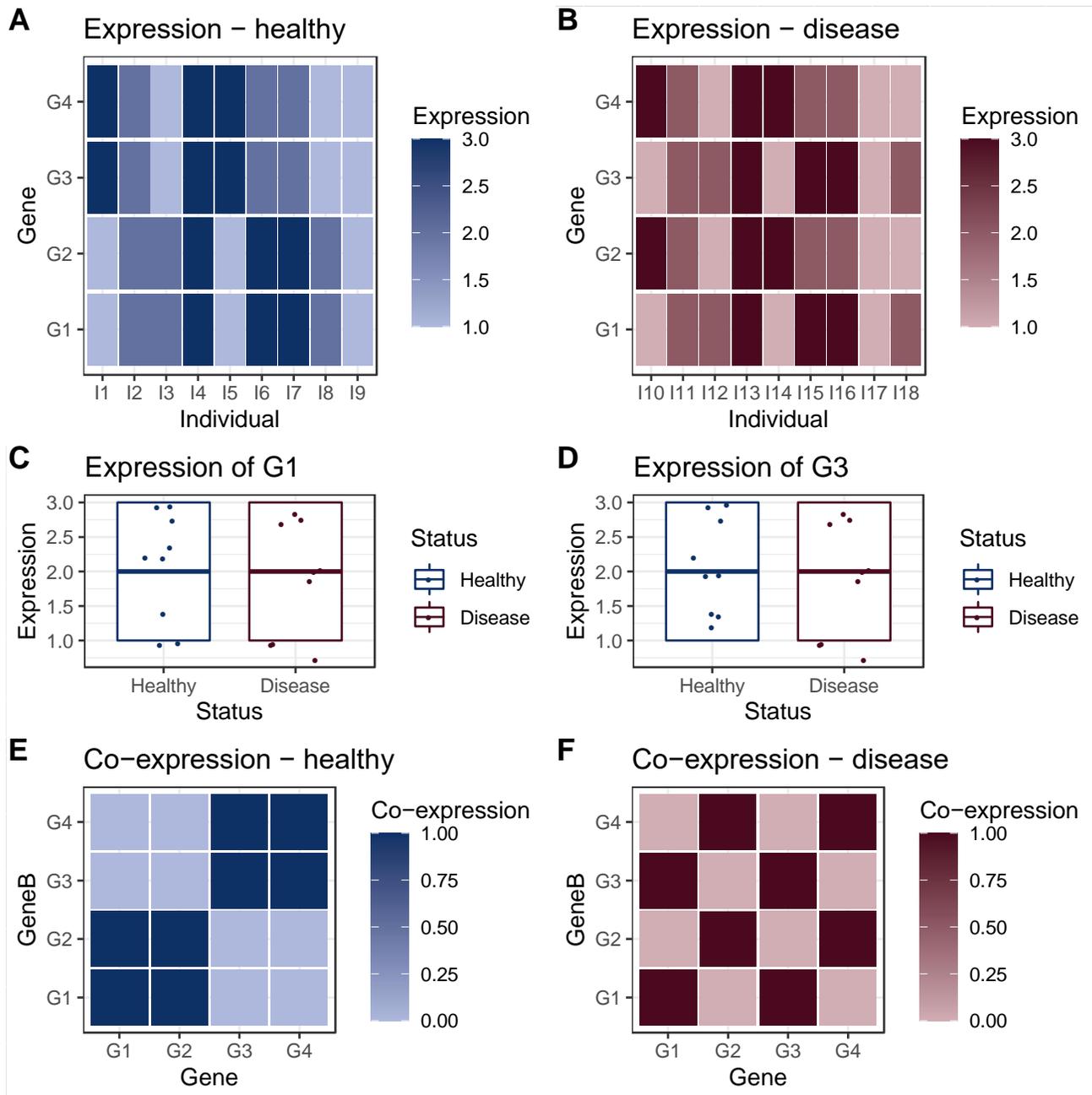

**Figure 1: Differential expression vs differential co-expression**. As a toy example, we consider the expression of four genes in (**A**) nine healthy individuals and (**B**) nine individuals with a disease. In this example, none of the genes are differentially expressed, as they have a similar average expression level in both healthy and disease individuals, as shown in examples (**C**) gene G1 and (**D**) gene G3. However, when we look at the co-expression between genes within healthy individuals (**E**) and within disease individuals (**D**), we see that there is obvious differential co-expression between genes in healthy individuals, compared to disease individuals.

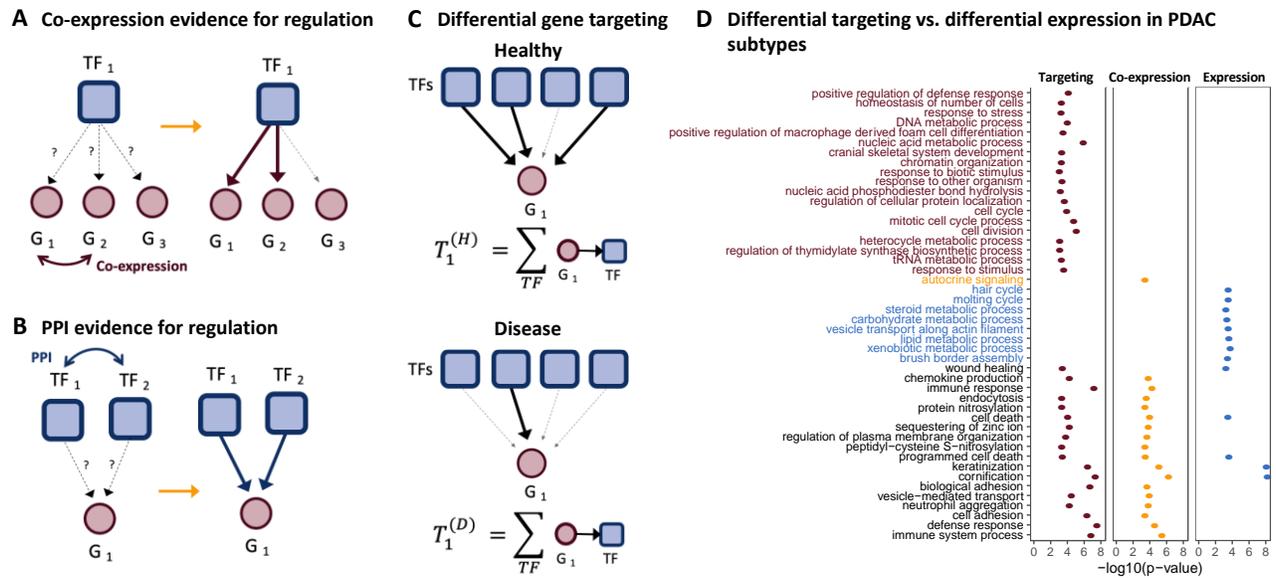

**Figure 2: Gene targeting.** Gene targeting scores are derived from GRNs and thus are influenced by the components used to derive the edge weights of a GRN. For example, PANDA GRNs include information regarding the **(A)** co-expression relationships between genes, and **(B)** protein-protein interactions between TFs. **(C)** Gene targeting scores are calculated as the sum of the weights across all of the inbound edges pointing to a gene. **(D)** GO enrichment of ranked differential gene scores comparing the basal-like and classical PDAC subtypes. Genes were ranked by differential targeting (red), differential co-expression (orange) and differential expression (blue).